\documentclass[%
 aip,
 jcp,
 amsmath,amssymb,
 reprint,%
]{revtex4-1}

\usepackage{xcolor}
\newcommand{\ie}{\emph{i.e.}}
\newcommand{\eg}{\emph{e.g.}}

\newcommand{\kBT}{{k_\te{B} T}}

\newcommand{\te}[1]{\mathrm{#1}}

\newcommand{\ev}[1]{\left\langle #1 \right\rangle}
\newcommand{\cb}[1]{\left\{ #1 \right\}}
\newcommand{\pt}[1]{\left( #1 \right)}
\newcommand{\sq}[1]{\left[ #1 \right]}

\usepackage{graphicx}% Include figure files
\usepackage{dcolumn}% Align table columns on decimal point
\usepackage{bm}% bold math
\usepackage[utf8]{inputenc}
\usepackage[T1]{fontenc}
\usepackage{mathptmx}
\usepackage{etoolbox}
\usepackage{blindtext}
\makeatletter
\def\@email#1#2{%
 \endgroup
 \patchcmd{\titleblock@produce}
  {\frontmatter@RRAPformat}
  {\frontmatter@RRAPformat{\produce@RRAP{*#1\href{mailto:#2}{#2}}}\frontmatter@RRAPformat}
  {}{}
}%
\makeatother
\begin{document}

\preprint{AIP/123-QED}

\title[Obstacle-tuned transitions in chiral active fluids]{Obstacle-tuned transition from chaotic to coherent vortex flows and odd diffusion in chiral active fluids}
% Force line breaks with \\

\author{Joscha Mecke}
 \affiliation{ 
Institute for Advanced Study, Shenzhen University, 518060 Shenzhen, P.R. China
}%
 \affiliation{ 
College of Physics and Optoelectronic Engineering, Shenzhen University, 518060 Shenzhen, P.R. China
}%
 \affiliation{ 
Theoretical Physics of Living Matter, Institute for Advanced Simulation, Forschungszentrum Jülich, 52425 Jülich, Germany
}%
\author{Yongxiang Gao}%
 \email{yongxiang.gao@szu.edu.cn}
\affiliation{ 
Institute for Advanced Study, Shenzhen University, 518060 Shenzhen, P.R. China
}%
\author{Marisol Ripoll}%
 \email{m.ripoll@fz-juelich.de}
\affiliation{ 
Theoretical Physics of Living Matter, Institute for Advanced Simulation, Forschungszentrum Jülich, 52425 Jülich, Germany
}%

\date{\today}% It is always \today, today,

\begin{abstract}
The interaction of a suspension of rotating colloids with a periodically patterned structure is here investigated by means of continuum theoretical predictions and hydrodynamic simulations. Close to the obstacle surface, rotors circulate opposite to their inherent direction of rotation as a result of unidirectional rotational stresses, which is in agreement with a prediction of the generalised Stokes equation. The resulting stationary background flow significantly affects the system dynamics and coexists with the intrinsic active turbulent behaviour. 
The relative importance of either of the two contributions can be controlled with the rotor density and the obstacle size, such that the system is either dominated by stationary vortices pinned to the obstacles or vivid active turbulent dynamics. %
While momentum dissipation into an underlying frictional substrate damps the related flows, small values of the friction can enhance the vortex flow around an obstacle. %
The colloids' diffusive dynamics are governed by odd diffusive fluxes guiding the colloids around the excluded volume introduced by obstacles, such that enhanced effective diffusive transport is obtained at finite obstruction. %
Our results pave the way to systematically address how confinement can be employed in order to control or harness the dynamics of colloidal chiral active turbulence and how the interplay of emerging edge currents and active turbulent dynamics at varying densities can be systematically determined.
\end{abstract}

\maketitle

%################################################%
\section{Introduction}
\label{sec:intro}
%################################################%

\noindent
Chiral active fluids consist of actively rotating building blocks, which can be externally actuated rotating colloids~\cite{katuri2024control}, or intrinsically active rotational swimmers~\cite{huang2021circular,tan2022odd}. 
The so-called circle swimmers are a prominent example characterised by the average individual circular trajectories~\cite{van2008dynamics, kummel2013circular,liebchen2022cam}, although most other isolated rotors do not necessarily display such circular trajectories. %
Colloidal chiral active suspensions contain rotating elements causing  a rotation of the surrounding solvent which induce the rotation of neighbouring particles. The particles' trajectories are then determined by the symmetry breaking of the local rotational friction leading to an unbalanced mutual rotational drive and the formation of multiscale vortices~\cite{mecke2023simultaneous}. 
Collective dynamics of such chiral active systems crucially depend on the rotor density, resulting into a Brownian low density limit, active turbulence at intermediate densities, and a slowdown of the dynamics induced by an increase of the effective solvent viscosity at very high colloidal densities~\cite{mecke2023simultaneous}. %
At larger scales, rotors can be realised as dry gears without any surrounding solvent, where the direct rotational contact friction causes a similar phenomenology~\cite{yang2020robust}. 

Geometries where confinement or interactions with obstacles are significant, are typically associated with hindrance of free diffusion and caging resulting from the obstruction within the complex geometries~\cite{bier2008self, zhou2020diffusion, ghosh2015non, polanowski2014simulation}. In active matter systems, the interaction of the active units with the solid walls can lead to rich effects such as locally pinned or directed dynamics, emergence of stationary vortex patterns, clogging, or even dynamics that show commensuration or frustration with the obstacle lattice~\cite{schimming2024vortex, nguyen2017clogging, reichhardt2021commensuration, reinken2020organising, reinken2022ising, nishiguchi2018engineering, doostmohammadi2019coherent, doostmohammadi2017onset, zhu2022colloidal}. % 
Boundaries and interfaces in chiral active fluids are known to excite edge fluxes~\cite{mecke2024emergent, dasbiswas2018topological, li2024robust}, leading to robust particle transport along the edges. The introduction of boundaries into odd diffusive systems leads to diffusive fluxes that constitute in a rolling effect along the boundary~\cite{kalz2022collisions}. %
In contact with an array of obstacles, circle swimmers exhibit either enhanced or diminished transport controlled by the obstacle density and noise~\cite{van2022role}. The enhancement can be explained by the fact that collisions between the active particles and the obstacles interrupt the swimmers limited circular trajectories, thus giving rise to trajectories with an enhanced effective diffusive behaviour~\cite{olsen2024optimal}. 
This raises the question whether boundaries can be used in chiral active matter systems in order to modify or even harness spontaneous flows.

In this article, we show that chiral active fluids consisting of a colloidal suspension of rotors in patterned environments, such as regular obstacle lattices, lead to a rich phenomenology such as tamed active turbulence, pinning of vortices around the obstacles, and enhanced odd diffusive transport. %%%
We derive the explicit solution of the Stokes flow featuring chiral activity~\cite{mecke2024emergent} around an isolated obstacle with hydrodynamic coupling between the no-slip obstacle and colloid surfaces. %
The predictions are tested with results obtained using massively parallel GPU accelerated explicit solvent multiparticle collision dynamics simulations of suspended colloidal rotors following an explicit experimental protocol~\cite{mecke2023simultaneous, mecke2024substrate}. %
Unbalanced rotational stresses lead to the formation of an edge current along the surfaces of the obstacles which increases with increasing rotor density and obstacle diameter. As a consequence of the viscous fluid forces, the edge current leads to the formation of a stationary background flow which coexists with the inherent active turbulent dynamics. The obstacles diminish the active turbulent dynamics to an extent that can be tuned by the rotor density, the obstacle size and separation, and also the friction of the substrate. Additionally, it is shown how the obstacles can be used to facilitate effective diffusive transport by guiding the rotors around the obstacles with odd diffusive fluxes. %

%################################################%
\section{Model}
\label{sec:model}
%################################################%
\noindent
We simulate a two-dimensional chiral active fluid composed of rotating colloids suspended to a solvent in a periodically patterned environment. %
For the solvent, we employ an angular momentum conserving variant of the well established mesoscopic hydrodynamic simulation method multiparticle collision dynamics (MPC)\cite{gompper2009multi}. %
Therein, the solvent consists of point-like fluid particles of mass $m$ whose positions $\bm{r}_i$ and velocites $\bm{v}_i$ are updated at discrete time steps of time difference $h$ according to the following two-step protocol. %
First, fluid particles move ballistically according to $\bm{r}_i(t+h) = \bm{r}_i(t) + \bm{v}_i(t)h$. %
Then, the particles are sorted into square collision boxes of length $a$ and the fluid particles within a respective box exchange momentum according to the following rule. The relative velocity of each fluid particle with respect to the mean velocity in the respective box is rotated by $\alpha = \pm \pi/2$ at equal probability. The final velocity of the fluid particle $i$ in collision box $\zeta$ after the collision at time $t+h$ then is composed of the average velocity in the collision box $\bm{v}_{\zeta}(t)$, plus the rotated relative velocity which yields $\bm{v}_i(t +h) = \bm{v}_\zeta(t) + \bm{R}(\alpha) \cdot (\bm{v}_i - \bm{v}_\zeta(t))$, where $\bm{R}(\alpha)$ is a two-dimensional rotation matrix. %
This protocol preserves linear momentum and energy. %
However, in order to prevent the emergence of unphysical torques~\cite{pooley2005kinetic}, we add a correction term which additionally enforces angular momentum conservation~\cite{noguchi2008transport, gotze2007relevance, mecke2023simultaneous}. %
Since we study an active matter system with constant energy input, we apply a thermostat to ensure an average constant system temperature $\kBT$. We employ the Maxwell-Boltzmann thermostat which ensures correct velocity rescaling on the level of the collision box~\cite{huang2010cell}. %
We employ a collision time of $h=0.02a\sqrt{m/(\kBT)}$ and an average number of fluid particles per collision box of $\rho = 10 m/a^2$, resulting in a solvent viscosity of $\eta = 17.9\sqrt{m\kBT}/a$, as accurately predicted using kinematic theory~\cite{noguchi2008transport}. In simulation units, we set $m = 1$, $\kBT = 1$, and $a = 1$. %

Colloids suspended to the solvent are modelled as movable no-slip boundaries that exchange linear and angular momentum with the fluid particles during streaming and collision steps, by applying the bounce back rule and virtual fluid particles in the colloid taking part in the collision step to optimise the no-slip behaviour~\cite{gotze2010mesoscale}. %
Colloids interact via a shifted repulsive WCA interaction corresponding to the potential
\begin{align}
\label{eq:potential}
    U(r) = \begin{cases}
        4 \epsilon\sq{\pt{\frac{a}{r-\sigma}}^{12} - \pt{\frac{a}{r-\sigma}}^{6}} + \epsilon\,, & \text{for }r \le \sigma + 2^{1/6}a\\
        0\,, & \text{else}
    \end{cases}
\end{align}
with $\epsilon = \kBT$, such that it is short-ranged and there is at least one collision cell in between the rotors in order to guarantee proper hydrodynamic coupling. %
The variables of the colloidal degrees of freedom are updated using a velocity-Verlet molecular dynamics scheme~\cite{gotze2011flow}. %
The colloids' rotational activity is implemented by ensuring a constant angular velocity on each of the colloids' surfaces and thus closely resembling experimental colloidal behaviour for rotational frequencies and magnetic field strengths where the colloids' magnetic moment precisely follow the rotation of the externally applied rotating magnetic field~\cite{mecke2023simultaneous}. As a result, a rotational flow field around each rotor is induced. %
We employ a fixed angular velocity of $\Omega = 0.01857/(a\sqrt{m/(\kBT)})$ or $\Omega = 0.01/(a\sqrt{m/(\kBT)})$. %
Considering the fluid flow created at the surface of a colloid, this results in Reynolds and P\'eclet numbers of $\mathrm{Re}=\sigma^2\Omega/(4\nu) \simeq 0.09$ and $\mathrm{Pe} = \sigma^2\Omega/(4D) \simeq 20$, respectively~\cite{mecke2023simultaneous}.

\begin{figure*}[ht!]
	\centering
  \includegraphics[width=1.\textwidth]{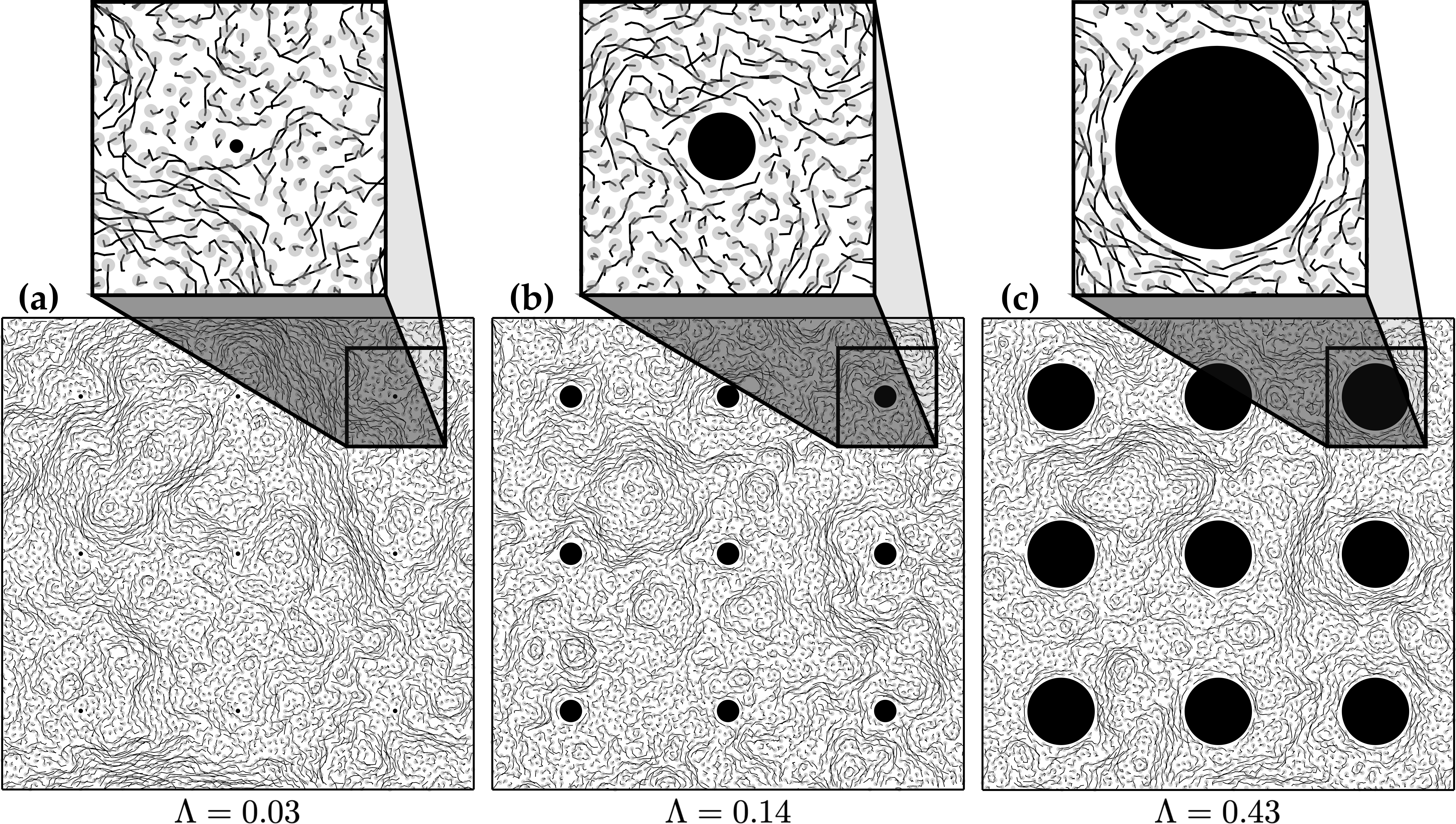}
	\caption{
        Constrained active turbulent eddy formation in the patterned environment. %
		Simulation snapshots and superimposed rotor trajectories for various lattice confinement parameters $\Lambda = D/L$, with varying obstacle sizes $D$, fixed separation between obstacles centers $L=35\sigma$, and  fixed rotors size $\sigma$, at a rotor density $\phi = 0.32$. %
        Zoom-in shows individual rotors' dynamics in the edge flow. %
		The trajectories are of duration $t\Omega/(2\pi) = 3$, \ie, the rotors perform three individual rotations along the trajectory. %
	}
	\label{fig:trajectories_obs}
\end{figure*}
We study the hydrodynamic chiral active fluid in a square periodic array of obstacles which we model by implementing non-movable no-slip walls. %
The rotor diameter is $\sigma = 6a$ and the diameter of the obstacle $D$ varies between $5\sigma$ and $160\sigma$ significantly exceeding the size of the rotors. The rotors interact with walls in a similar fashion as in Eq.~\eqref{eq:potential}, \ie, a 12-6 shifted WCA potential ensures that the minimal distance between the surface of the rotors and the wall is at least one collision box. %
Despite the fact that we are only analysing the rotor degrees of freedom here, the main computational effort is expended for the dynamics of the solvent particles. We have developed a highly parallelised GPU based simulation code running on high-end GPU-supercomputers~\cite{kesselheim2021juwels} in order to be able to simulate the low-Reynolds number dynamics of up to $10^5$ colloids and $10^7$ fluid particles. %

%################################################%
\section{Results}
\label{sec:results}
%################################################%
%################################################%
\subsection{Superposition of active turbulent and directed dynamics}
\label{sec:superposition}
%################################################%
\noindent
A patterned environment is considered by placing fixed circular obstacles in a two-dimensional periodic square lattice. 
The confinement exerted by the lattice can be characterised by the ratio between the obstacle diameter $D$ and the distance between obstacle centres $L$, this is $\Lambda \equiv D/L$. 
With this definition, the limit $\Lambda \to 0$ corresponds to a bulk system with no obstacles, while $\Lambda \to 1$ corresponds to a system with obstacles in contact and completely separated domains in-between each four obstacles. We refer therefore to $\Lambda$ as the lattice confinement parameter. %
In the unconfined limit $\Lambda \to 0$, the bulk dynamics is recovered, where the simultaneous emergence of active turbulence and odd dynamics has been described~\cite{mecke2023simultaneous}.  
Each rotor induces the rotation of the solvent together with its surface, such that all neighbouring rotors get a propulsion thrust, making them to rotate not only around their axes but also around each other. In the presence of a large number of rotors, this leads to the emergence of eddies of different sizes, together with an accumulation of the rotors in areas rotating in the same direction as the rotors, and a depletion in counter-rotating areas.

With increasing $\Lambda$, this is with increasing size of the obstacles, the instantaneous dynamics alters in two main manners (see Supplementary Movie 1).
On the one hand, due to the obstacles excluded volume, the formation of vortices is hindered by the excluded volume defined by the obstacles, leading to a decreasing vortex size. %
One the other hand, the rotors tend to displace along the obstacles surface, such that an effective edge flow around the obstacles emerges. Both effects become more important with increasing obstacle size. An increasing rotor density also enhances the effect because the mutual steric hindrance of the rotors crucially limits their free translation and thus the formation of vortices. 
Simulations with nine obstacles and periodic boundary conditions are performed for systems with rotors density $\phi=0.32$ and various values of $\Lambda$. Short time rotors trajectories in Fig.~\ref{fig:trajectories_obs} show the instantaneous formation of unordered vortices of various sizes, and it can be observed that the maximum vortex size decreases with increasing obstacle size. The appearance of flow along the obstacles surface is also more significant the larger the obstacles are.

\begin{figure*}[t]
    \includegraphics[width=1.\textwidth]{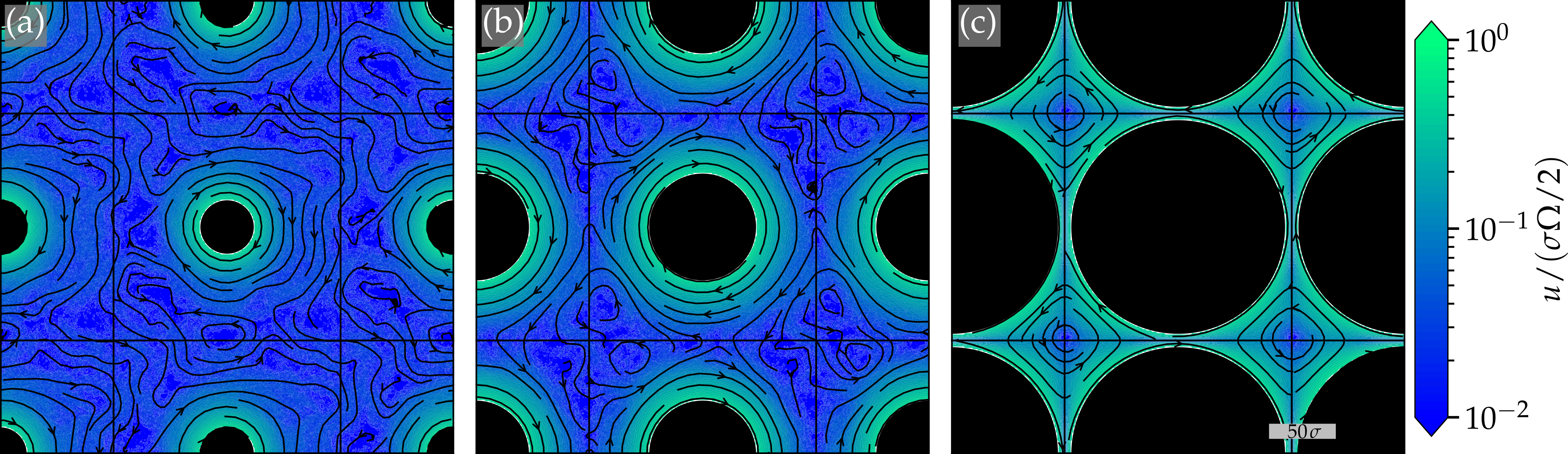}
	\caption{
        Rotor stream lines for simulations with one single unit cell and different confinements: (a)~$\Lambda = 0.24$, (b)~$\Lambda = 0.47$ and (c) $\Lambda = 0.94$. Colour map indicates the velocity magnitude $u = |\bm{v}|$ showing how the edge flows propagate to the interior of the system.  The data are obtained  from 640 independent simulations for $\phi = 0.32$ with $L=170\sigma$ and the normalization factor $\sigma \Omega / 2$ corresponds to the rotor surface velocity. %
	}
	\label{fig:flow-field}
\end{figure*}
The flow induced along the edges becomes more obvious when the rotors velocity is averaged over large time intervals. The velocity field and corresponding stream lines are shown in Fig.~\ref{fig:flow-field} where, in order to increase efficiency, simulations with one single unit cell are used. 
Close to the obstacles surface, the flow rotating around the obstacles in the opposite to the colloids rotation becomes obvious. For increasing distance, the flow decays and thermal fluctuations become more apparent. %
The created flows at the edges of the unit cell vanish due to continuity and the periodicity of the system, 
giving rise to a star shaped orbital counter rotating flow field within the area between four obstacles.
For smaller values of $\Lambda$, the directed flow is weak and fluctuations dominate as in Fig.~\ref{fig:flow-field}a, for larger $\Lambda$ in Fig.~\ref{fig:flow-field}b, the combination of edge flow and counter-rotating areas becomes more obvious, and for the largest confinement case $\Lambda \to 1$ in Fig.~\ref{fig:flow-field}c, the flow essentially reduces to a single vortex in between the obstacles.    
%

%################################################%
\subsection{Edge flow quantification}
%################################################%
\label{sec:directed}
%################################################%
\noindent
An analytical estimation of the edge flow can be performed by considering the colloidal chiral active fluid as a two-dimensional continuum under the influence of viscous and active stresses together with internal pressure. 
The coarse-grained flow dynamics $\bm{v}(\bm{r})$ of $N$ rotors with angular velocity $\Omega_i$ at position $\bm{r}_i$, and a corresponding local angular velocity density $\widetilde\Omega=\ev{\sum_i\Omega_i \delta(\bm{r}-\bm{r}_i)}$ 
can then be described by the generalised incompressible Stokes equation~\cite{mecke2024emergent}, 
\begin{equation}
\label{eq:incom-stokes}
     -\frac{1}{\rho} \nabla p_\te{eff} + \frac{\eta}{\rho} \nabla^2 \bm{v} + \frac{\eta_\te{R}}{\rho}\hat{e}_z \times \nabla (2\widetilde\Omega - \omega) = 0\,.
\end{equation}
Here, $\rho$ is the constant fluid density, %$\widetilde\Omega=\ev{\sum_i\Omega_i \delta(\bm{r}-\bm{r}_i)}$ is the local angular velocity density, 
and consequently the angular velocity density can be assumed to be constant $\widetilde\Omega \propto \Omega \rho$. Furthermore, $p_\te{eff}\equiv p - \eta^\te{odd} \omega$ is an effective pressure that balances the bulk influences of odd viscosity and ensures incompressibility together with the continuity equation $\nabla \cdot \bm{v} = 0$ and the vorticity is defined as $\omega = \hat{e}_z\cdot(\nabla\times\bm{v})$. %
The term proportional to the shear viscosity $\eta$ is the usual Laplacian shear, %
while the term proportional to the rotational viscosity $\eta_\te{R}$ acts as a synchronisation of the individual colloids rotation and the rotor fluid vorticity and thus couples the inherent rotation of the rotors to the translational degrees of freedom. %

\begin{figure*}[t]
    \includegraphics[width=1.\textwidth]{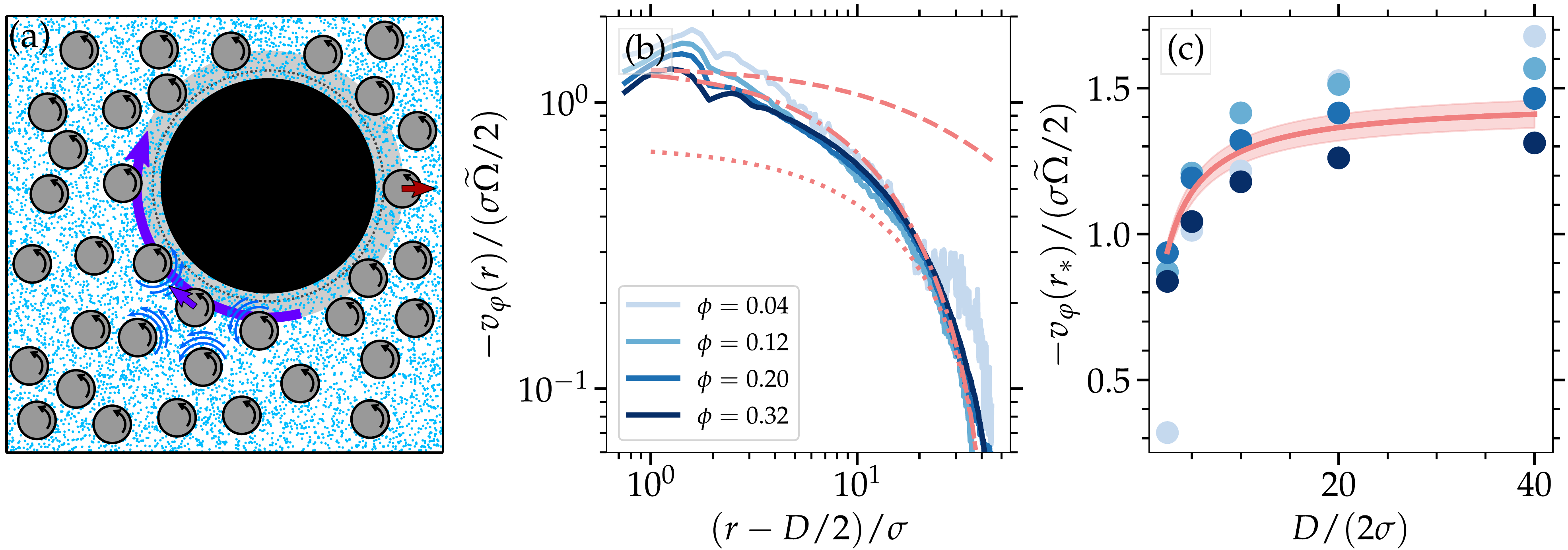}
	\caption{%
		Edge flows and a stationary flow profile. %
        (a)~System sketch with fixed obstacle in black, rotating colloids in grey, and 
        solvent particles in blue. Emergence of unidirectional hydrodynamic stresses at the boundary leading to the edge flow illustrated by blue arrows, steric wall-colloid interactions by the red arrow and excluded volume by a shaded area. %
        (b)~Azimuthal flow profile around an obstacle with $D/\sigma = 80$ normalised by angular velocity density $\widetilde\Omega$. The dashed line corresponds to Eq.~\eqref{eq:obs-flow}, 
        the dotted line corresponds to the same line displaced by the value at the unit cell boundary, $v_\varphi[(L-D)/2]$, 
        and  the dash-dotted line corresponds to a quadratic interpolation between the two previous ones. 
        (c)~Normalised average colloid velocity directly at the obstacle boundary $r_\ast = D/2 + 1.25\sigma$ against size of the obstacle at different rotor densities $\phi$. Symbols correspond to simulation data and the line to a least-square fit to the points according to Eq.~\eqref{eq:obs-flow}. Standard deviation obtained from the fit depicted as shaded area.
        The data in (b) and (c) are obtained from five simulations of length $T\Omega/(2\pi) = 2950$ and four simulations of $T\Omega/(2\pi) = 1030$, with one single unit cell, and $L=170\sigma$.
	}
	\label{fig:edge-flow}
\end{figure*}

The flow around a single isolated obstacle is first considered.  
The steady-state solution for the rotors flow field in the azimuthal direction is enough to describe the relevant system dynamics given the rotational symmetry, this is $v_\varphi(r)$ with $r$ the distance to the obstacle centre, a scenario which is geometrically related to the flow within a circular container~\cite{liu2020oscillating}. The pressure cancels out of the equation in the azimuthal direction, such that Eq.~\eqref{eq:incom-stokes} reads in polar coordinates ($r$, $\varphi$) %
\begin{align}
    0 = (\eta + \eta_\te{R}) \cb{\partial_r^2 v_\varphi + \frac{1}{r}\partial_r v_\varphi - \frac{v_\varphi}{r^2}}\,.
\end{align}
The solution takes the form $v_\varphi(r) = c_1 r + c_2/r$, where the constants $c_1$ and $c_2$ are to be determined using the boundary conditions. For an isolated obstacle, we assume $v_\varphi(r \to \infty) = 0$, thus fixing $c_1=0$. %
%At the obstacle surface, the colloidal rotors do not experience stress directly through the boundary, but the solvent is subject to a no-slip boundary condition on the surface of the obstacle and the rotors, where the minimal distance between rotor and obstacle surfaces is $a$. We assume now that t
The boundary condition at the obstacle surface is a little more subtle, because the colloidal rotors do not experience stress directly through the boundary, but by the viscous coupling to the solvent which is subject to a no-slip boundary condition on the surface of the obstacle and the rotors. %
To overcome this problem, a first layer of rotors can be considered to be evenly distributed with density $\rho$ at an effective distance $\delta$ and with a persistent rotation due to the emerging edge current. %
The viscous stress between the obstacle and the first rotor layer can be approximated by the viscous stress between two coaxial cylinders with a narrow gap $\delta$ in between. %, see Fig.~\ref{fig:edge-flow}a.
For small $\delta$ values, this stress has been calculated as~\cite{landau1987fluid} $\Sigma_\delta = \eta v_\varphi(D/2)/\delta$, and here we take $\delta = \sigma/2 + 2^{1/6}a$, with $a$ the minimum allowed distance between rotor and obstacle surface.  %
The internal stress tensor of the chiral active fluid consists of shear~\cite{landau1987fluid} and rotational stresses~\cite{banerjee2017odd} 
\begin{equation}
    \Sigma_{\varphi r} = \eta \cb{\partial_r v_\varphi - \frac{v_\varphi}{r}} + \eta_\te{R} \cb{\frac{v_\varphi}{r} + \partial_r v_\varphi - 2\widetilde\Omega}.
\end{equation}
Inserting the solution $v_\varphi = c_2/r$ into $\Sigma_{r\varphi}\vert_{r=D/2} = \Sigma_\delta$ enables us to find the solution of the flow around an isolated obstacle,
\begin{align}
\label{eq:obs-flow}
    v_\varphi(r) = -\frac{\eta_\te{R}}{\eta} \frac{\widetilde\Omega D^2}{D/\delta + 4} \frac{1}{r}\,.
\end{align}
This solution agrees with the observed collective circulation of the rotor fluid opposite to the intrinsic individual rotors rotation in Fig.~\ref{fig:flow-field}. The dominant contribution to collective rotations in rotor fluids stems from the breaking of symmetry and thus the unidirectional rotational stress experienced by the rotor layers at the obstacle boundary as illustrated in Fig.~\ref{fig:edge-flow}a. %
Furthermore, $v_\varphi$ is directly proportional to $\widetilde{\Omega}$ (and thus also proportional to the rotor density) and $\eta_\te{R}$ resulting from the internal driving of the rotors and the ability of the solvent to couple the rotors' internal rotation to the circulation of the rotor fluid around the obstacle. %

Velocity profiles for the rotors fluid are shown in Fig.~\ref{fig:edge-flow}b for simulations in a periodic domain at different rotor densities with $v_\varphi$ averaged over the whole range of azimuthal angles. 
The data collapse obtained with the normalisation of the profiles with  $\widetilde\Omega$, emphasizes the direct proportionality of  $v_\varphi$ with both $\Omega$ and $\rho$, as predicted by Eq.~(\ref{eq:obs-flow}).
For small values of $r/\sigma$, this is, close to the obstacle boundary an oscillatory velocity profile appears which has been previously reported in chiral active systems~\cite{liu2020oscillating}. These oscillations quickly decay with distance from the wall, showing the fluid on average can be accurately described by a fluid of constant density as in Eq.~\eqref{eq:obs-flow}. %
Also close to the obstacle boundary, simulations with very low densities do not exactly collapse with the others since the number of colloids is low and the rotor fluid density can barely be approximated as constant. 
On the other hand, far from the obstacle surface the flow decays faster than predicted by Eq.~(\ref{eq:obs-flow}). This is due to the superposition of the flow velocities of the neighbouring obstacles which have opposite directions and ensure that, on average, the flow necessarily vanishes at the unit cell boundary.
Accounting for the far flow behaviour can be done by simply subtracting the value of $v_\varphi$ at the boundary ($r=L/2$) from Eq.~\eqref{eq:obs-flow} in order to force the flow to vanish at $r=L/2$ (dotted line in Fig.~\ref{fig:edge-flow}b). 
A quadratic interpolation scheme between the two lines matches the simulated profile, showing that we indeed obtain a crossover defined by the flow dictated by the edge current and continuity across the periodic image. %

In order to evaluate the intensity of the flow as a function of the density and obstacle size, we compare the velocity at fixed distance very close to the obstacle surface, $r_\ast= D/2 + 1.25\sigma$.
After plugging in $r_\ast$ into Eq.~\eqref{eq:obs-flow}, we note that the edge flow attains a constant for $D\to\infty$.
Comparison to simulation results are shown in Fig.~\ref{fig:edge-flow}c together with Eq.~\eqref{eq:obs-flow} for $v_\varphi(r_\ast)$ as a function of $D/(2\sigma)$, where the value of $\eta_\te{R}/\eta = 0.53 \pm 0.02$ is obtained as a best fit (also used in Fig.~\ref{fig:edge-flow}b).
Note that the value of $\eta_\te{R}$ in Eq.~\eqref{eq:obs-flow} is not \emph{a priori} known for the MPC solvent, such that this fit to the simulation data can be considered as an indirect measurement of $\eta_\te{R}$.  %
The obtained agreement is  very satisfactory. The deviations are density dependent and therefore consistent with an undetermined weak density dependence of both $\eta$ and $\eta_\te{R}$~\cite{haines2008effective}. 
Note, that in our treatment the edge flow is solely initiated by the unidirectional rotational stresses exerted by neighbouring rotors directly at the wall. Hydrodynamic forces between a rotating colloid and a convex no-slip boundary are not taken into account since they are expected to be negligible. 

\begin{figure*}[ht!]
	\centering
    \includegraphics[width=1.\textwidth]{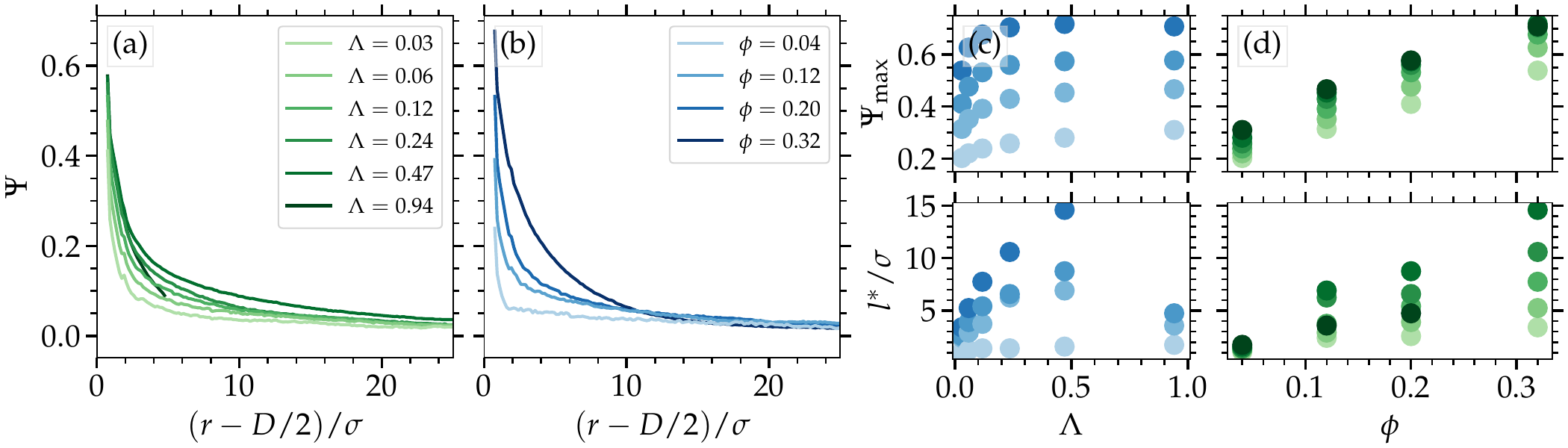}
	\caption{
		Transition from chaotic to ordered circular trajectories around the obstacles. %
        (a)~Vortex order parameter at $\phi=0.20$ for varying obstacle size $\Lambda$ revealing that the circular flow around the obstacle is maximised for large obstacles. %
        (b)~Vortex order parameter for $\Lambda=0.12$ for varying rotor density $\phi$ displaying an increase of circular flow around the obstacle with increasing density. %
        Both trends in (a) and (b) are captured in (c) and (d) showing in the top panel the height of the dominant maximum at the obstacle fluid interface $\Psi_\te{max}$ and in the bottom panel, the penetration depth of the vortical order $l^\ast$ which satisfies $\Psi(l^\ast) = 0.1$, against $\Lambda$ (c) and $\phi$ (d). %
        Statistics as in Fig.~\ref{fig:edge-flow}. %
	}
	\label{fig:vortex_order}
\end{figure*}

\subsection{Pinning vortices to obstacles}
\noindent
The actual particle trajectories in Fig.~\ref{fig:trajectories_obs} are largely chaotic, while the edge velocities characterised until now in Fig.~\ref{fig:flow-field} and Fig.~\ref{fig:edge-flow} refer to an average over large times and different simulations. In order to quantify how strongly pinned are the individual rotor trajectories around the circular obstacles we define the {\em pinned vortex order parameter} as~\cite{wioland2013confinement, lushi2014fluid}
\begin{align}
\label{eq:vortex-order}
    \Psi(r) = \frac{1}{1-\frac{2}{\pi}}\left[\frac{\ev{|\bm{v}_i \cdot \hat{e}_\varphi(\bm{r}_i)|}_{r_i\in S(r)}}{\ev{|\bm{v}_i| }_{r_i\in S(r)}} - \frac{2}{\pi}\right],
\end{align}
where $\hat{e}_\varphi$ is the unit vector into the azimuthal direction, $\ev{\cdot}_{r_i\in S(r)}$ denotes an average over all rotors $i$ with positions $\bm{r}_i$ in a circular shell of width $\Delta r = \sigma/6$ at radius $r$, and the rotors' positions are measured with respect to the centre of the obstacle. %
The value of $\Psi(r)$ is then averaged over trajectories and realizations. %
For perfect persistently circular, carousel-type, trajectories around the obstacle, $|\bm{v}_i \cdot \hat{e}_\varphi(\bm{r}_i)| = |v_i|$ and thus $\Psi = 1$. %
For a completely random trajectory, $\bm{v}_i$ and $\hat{e}_\varphi(\bm{r})_i$ are  decorrelated, 
and $\ev{|\bm{v}_i \cdot \hat{e}_\varphi(\bm{r}_i)|} = \ev{|\bm{v}_i|} \ev{|\cos\vartheta|} = 2\ev{|\bm{v}_i|}/\pi$,  such that $\Psi = 0$. %
For radial trajectories, we have $\Psi < 0$. %
In the bulk case, the absence of obstacles makes the system translational invariant which corresponds precisely to $\Psi = 0$.
The parameter can thus be regarded as an order parameter for vortical dynamics pinned around the obstacles with respect to the radial distance from the surface. %
The colloids' instantaneous velocity bears a strong influence of thermal dynamics due to the collisions of the colloids with the fluid particles. These contributions lead to a systematic decrease of $\Psi(r)$, because the thermal contributions are uncorrelated and yield $\Psi(r)=0$. 
Accordingly, we use pre-averaged Euler rotor velocities for the calculation of $\Psi(r)$ in Eq.~\eqref{eq:vortex-order}, \ie~$\left[\bm{r}_i(t+\Delta t) - \bm{r}_i(t)\right]/\Delta t$, with~$\Delta t\Omega = 0.4$, the time in which the rotors have travelled a distance of about~$\sigma/6$. 

The obtained values for $\Psi(r)$ are displayed for varying confinement parameters in Fig.~\ref{fig:vortex_order}a and varying density in Fig.~\ref{fig:vortex_order}b. %
The rotors' dynamics show circular trajectories predominantly at the obstacle surface which is related with $\Psi_\te{max}$, the maximum value of $\Psi$ at the smallest $r$ values. 
With increasing distance to the obstacle, the rotors loose their coherence with the edge flow such that the trajectories become less circular and $\Psi$ decays with $r$. %
This decay can be characterized by the penetration depth $l^\ast$, this is the value at which $\Psi(l^\ast) \equiv 0.1$. The dependences of $\Psi_\te{max}$ and $l^\ast$ with both $\Lambda$ and $\phi$ are shown in Fig.~\ref{fig:vortex_order}c,~d. 
Larger values of $\Lambda$ and of $\phi$ show larger values of $\Psi_\te{max}$, with a similar dependence to the rotor velocity at the edge, as shown in Fig.~\ref{fig:edge-flow}c.
This similarity is due to the fact that the edge current leads to the vortical order $\Psi$, although
the values for $|v_\varphi(r)|$ and $\Psi(r)$ are not directly proportional, since deviations from the steady-state profile in $v_\varphi$ are averaged out, while in $\Psi(r)$ the instantaneous deviations lead to a systematic smaller value. 
The penetration depth $l^\ast$ increases monotonically with both $\Lambda$ and $\phi$, stating that the circular flow around the obstacle has a deeper impact on the created flows in the chiral active fluid around the obstacle when the edge current is stronger. Additionally, at large densities $\phi$, the active turbulent vortex formation is prevented by steric interaction between the rotor~\cite{mecke2023simultaneous}, such that the contributions of the edge current to the overall dynamics become more important. %
For $\Lambda \to 1$ the surfaces of neighbouring obstacles are increasingly close to each other and the penetration length saturates, for $\Lambda = 0.94$ for example $l^\ast$  reaches a maximum of $5\sigma$ which is exactly in between both obstacles.

\begin{figure*}[t]
    \includegraphics[width=1.\textwidth]{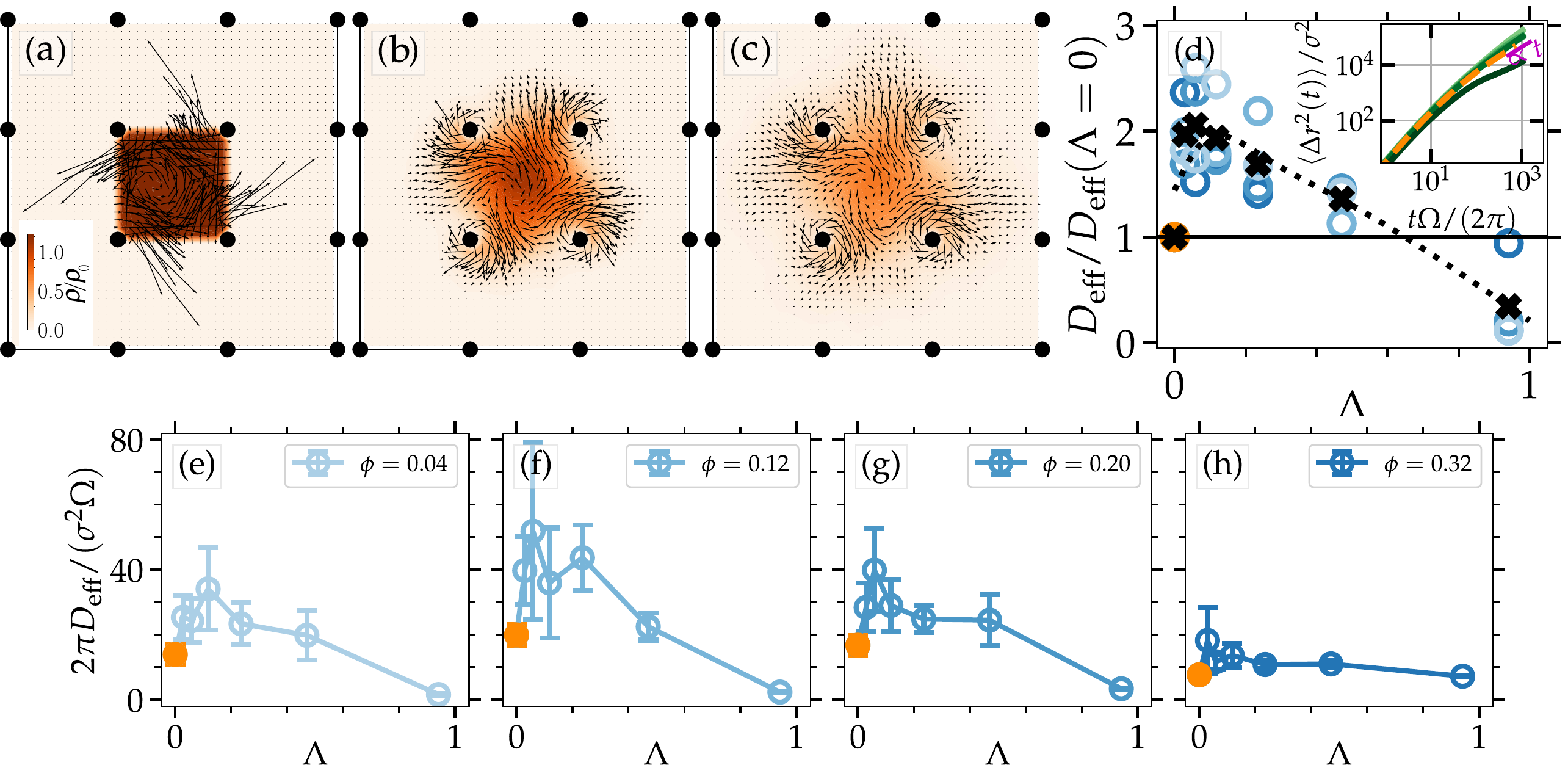}
	\caption{
        Effective diffusive chiral mass transport in an obstacle lattice. %
		(a)-(c)~Evolution of tagged rotor density in a system of constant density. 
        Time increases from left to right $t\Omega/(2\pi) = 0, 222, 444$, background color code indicates density of tagged rotors and the arrows the direction and intensity of local particle flux. 
        The non-perpendicular angle of particle flux with the density gradient indicates that the dynamics is translationally non-invariant and that both parallel and perpendicular components of the diffusion are non-vanishing.
        Simulation parameters are $\phi=0.32$, $L=35\sigma$, and $\Lambda=0.14$. 
        (d)~Effective diffusion coefficient normalised with bulk value $D_\te{eff}(\Lambda=0)$ (orange bullet) versus $\Lambda$ at different densities, colour code as in Fig.~\ref{fig:vortex_order}b. %
        Black crosses denote average at given $\Lambda$ over all densities and black dotted line represents a corresponding spline interpolation. %
        Inset: Mean-square displacement of the rotors for varying obstacle size $\Lambda$ at fixed density $\phi=0.20$, colour code as in Fig.~\ref{fig:vortex_order}a. Terminal diffusive regime indicated by magenta line. Orange dashed line indicates bulk results, \ie, $\Lambda = 0$. %
        (e)-(h)~Effective diffusion coefficient for varying density normalised by area exploration rate in units of square-diameters per rotor circulation. All lines shown to same scale. %
        Orange bullets correspond to a bulk system without obstacles. Simulation domain size in (d)-(h) $L=170$. %
	}
	\label{fig:tracer-diffusion}
\end{figure*}

\subsection{Odd and enhanced effective diffusion in a patterned environment}
\noindent
Isolated rotors move with a purely Brownian motion with a constant diffusion coefficient $D_0$, which is identical to that of a passive colloid. 
In the presence of neighbouring rotors, self-diffusion is enhanced since particles mutually roll over one another's surface and thus escape the phase-space-limiting effect of mutual steric hindrance upon inter-particle collisions. %
The induced translational motion of the rotors enlarges the standard parallel self-diffusion coefficient $D_\parallel$, \ie, parallel to the density gradient, while the transversal interactions lead to the emergence of fluxes perpendicular to density gradients, which manifest in odd or anti-symmetric contributions to the diffusion tensor, this is the odd diffusion coefficient $D_\perp$~\cite{mecke2024emergent, hargus2021odd, hargus2025flux}, typical of chiral active materials. %
Fick's law for the connection of diffusive flux $\bm{j}$ and the colloidal density $\rho$ reads $\bm{j} = - \mathbb{D} \cdot \nabla \rho$ with the diffusion tensor $\mathbb{D}$ being composed of self-diffusion coefficient $D_\parallel$ in the diagonal and the odd diffusion coefficient $D_\perp$ as anti-symmetric elements. 
In bulk conditions, the divergence of the anti-symmetric tensor elements vanish and only the even diagonal elements remain such that the diffusion equation remains unaltered. %
However, the transverse interactions are the microscopic origin for the rotors translational velocity which then increase with density~\cite{mecke2023simultaneous} and consequently $D_\parallel$ increases also with $\phi$. An increase of $\phi$ leads at the same time to an increase of the effective solvent viscosity as interparticle collisions become more frequent, ultimately slowing down particle transport. %, upon further increase of $\phi$. %
An optimal trade-off between transport facilitation and hindrance at $\phi \sim 0.12$ has been already investigated~\cite{mecke2023simultaneous} from the study of the mean-square displacements. %
The odd diffusion coefficient can also be measured in bulk systems making use of Green-Kubo-like relations~\cite{hargus2021odd} $D_\perp = \lim_{t\to\infty}(\ev{\Delta r_y(t) v_x(0)} - \ev{\Delta r_x(t) v_y(0)})/2$ to yield $D_\perp = (14\pm10) D_0$, measured from twelve individual simulations of total length $T\Omega/(2\pi)=1,480$ at $\phi=0.32$, where $D_0$ is the diffusion coefficient of isolated rotors, which is identical to the diffusion coefficient of a passive colloid~\cite{mecke2023simultaneous}.

The introduction of obstacles interacting with the chiral active fluid modifies the diffusive density evolution~\cite{kalz2022collisions}.
At the obstacle boundaries $B$, the no-flux condition, \ie, $(\mathbb{D}\cdot\nabla\rho)\cdot\bm{n}\vert_{B}=0$, with $\bm{n}$ the normal vector to the surface, only limits the diffusive flux normal to the obstacle surface but not the corresponding diffusive fluxes perpendicular to the surface stemming from the perpendicular fluxes.
% normal introduced by the chiral activity, such 
This means that the rotors do not only roll around one another~\cite{kalz2022collisions,langer2024dance}, but also along the obstacle surface, inducing a directional diffusive flux, and an overall enhancement of the diffusion due to the presence of the obstacle, which we denote as $D_\te{eff}$. %
To illustrate this extent, Fig.~\ref{fig:tracer-diffusion}a-c shows the time evolution of the thus obtained non-homogeneous diffusive dynamics of tagged rotors (see also Supplementary Movie 2). 
We are following the dynamics of tagged rotors in a system of approximately homogeneous density, the overall density remains constant.
We quantify the time evolution of the density of such tagged rotors, as well as their local flux. Density is calculated in square bins of area $(2.5\sigma)^2$, and the flux is obtained by taking into account how many individual tagged rotors move into neighbouring bins per time and line length $2.5\sigma$. Both density and flux are time averaged over the full trajectory, over nine equivalent areas in a simulation domain of nine obstacles, and over twelve simulations in order to reduce fluctuations. %
The tagged particles spread out with a clear directionality due to the presence of the obstacles as shown in Fig.~\ref{fig:tracer-diffusion}. The tracer density cloud entwines around the walls of the obstacles and a directed flow along the walls is created, leading to the active transport of rotors to neighbouring unit cells of the lattice. %
The emergence of flux densities not only down, but perpendicular to the tracer density gradient and along the obstacle surfaces into the direction predetermined by the chirality axis is a clear sign of odd diffusive contributions. %
Note, that although the introduction of obstacles and boundaries into the chiral active fluid leads to a drastic change of the overall dynamics, the origin of the transverse interactions and thus $D_\perp$ remains unaffected. 

In order to characterise the effect of the obstacles in the overall diffusion, we measure the parallel diffusion coefficient as obtained from the long-time limit of the time and ensemble averaged rotors' mean-square displacement (MSD, Fig.~\ref{fig:tracer-diffusion}d inset) in the obstacle lattice $\langle \Delta r^2(t) \rangle \underset{t\to \infty}{\rightarrow}4D_\te{eff}t$. %
The measured diffusion coefficients are plotted as a function of the  lattice confinement parameter $\Lambda$ as is shown in Fig.\ref{fig:tracer-diffusion}d for different colloidal densities $\phi$. 
The diffusion coefficient in bulk, $D_\te{eff}(\Lambda \to 0) = D_\parallel$, this is of a system without obstacles at the respective density, is used as a normalisation factor, and the measured values are presented together with the average over all densities at given $\Lambda$. %
We can clearly see that the pattern of the environment systematically enhances the effective diffusive particle transport over a broad range of $\Lambda$. %
This diffusion enhancement is obvious for small values of $\Lambda$ and reaches a maximum at $\Lambda \simeq 0.1$ with an average enhancement of about a factor of $2$, with respect to a bulk system ($\Lambda = 0$). %$D_\te{eff}$ is only smaller than in a system without pattern only for $\Lambda > 0.5$. %
Only after $\Lambda$ exceeds $0.5$, the effective constrain or cage by the obstacles cannot be compensated by the transport facilitation by ``rolling around'' the obstacle wall, such that the effective diffusion coefficient becomes smaller in comparison to a system without obstacles. %
Evidently, for $\Lambda \to 1$ the passage connecting the two areas enclosed by four obstacles each becomes very narrow, such that the MSD is indicative of temporal caging, \ie, it shows to saturate and becomes diffusive again at late times (see inset of Fig.~\ref{fig:tracer-diffusion}d). %
For $\phi = 0.32$ and $\Lambda = 0.96$, the created fluxes at the obstacles' boundaries dominate the overall diffusive dynamics, similar to the behaviour of the edge flow and the vortex pinning. As a result, the enhancement and decrease of the effective diffusive transport by virtue of boundary fluxes and excluded volume, respectively, cancel, such that $D_\te{eff}$ approaches the value of a system without boundaries at the same density. This is in contrast to systems at lower density, where the $D_\te{eff}$ is dominated by transport obstruction by excluded volume. %

The data in Fig.~\ref{fig:tracer-diffusion}d is individually shown in Fig.~\ref{fig:tracer-diffusion}e-h to make explicit the dependence of the effective diffusion coefficient normalised by the colloidal space exploration per circulation $2\pi D_\te{eff}/(\sigma^2\Omega)$ versus $\Lambda$ at each density. % individually. %
Additionally, the density dependence of the effective diffusive transport in the obstacle lattice can be inferred here, showing that $D_\te{eff}$ first increases with density, reaches a maximum at about $\phi = 0.12$, and decreases upon further increasing the density, in accordance to the behaviour in a system without obstacles~\cite{mecke2023simultaneous}. %

\begin{figure*}[t]
    \includegraphics[width=1.\textwidth]{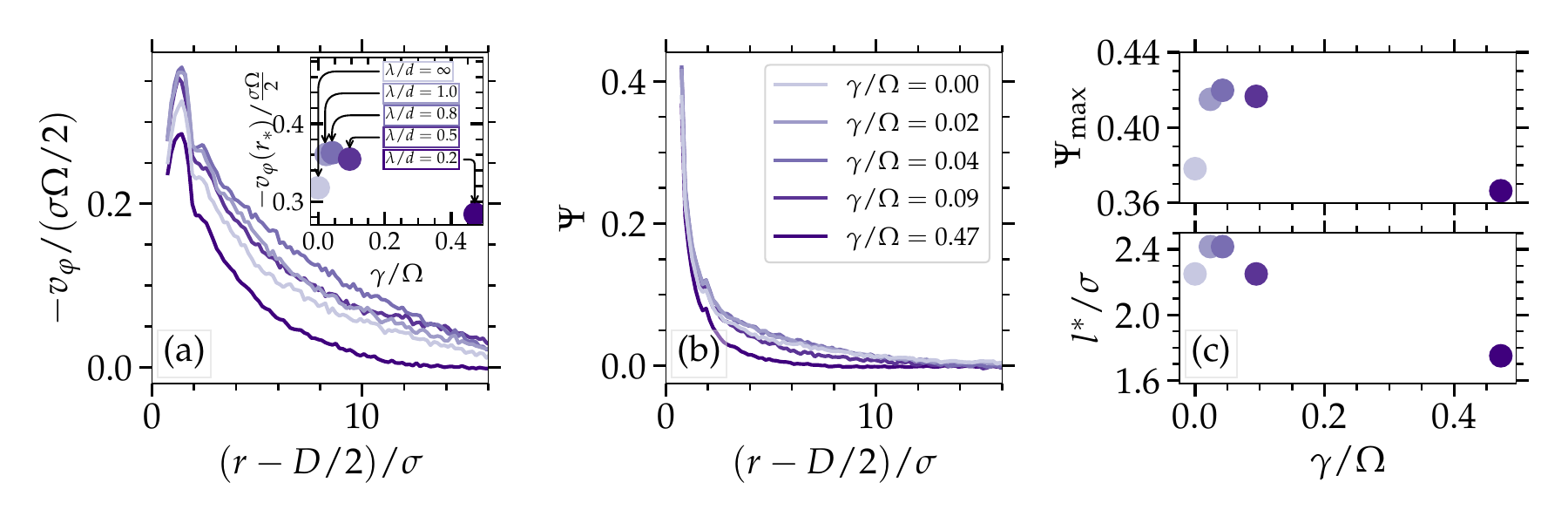}
	\caption{
        Vortex order under the influence of a frictious substrate.
        (a)~Azimuthal velocity profile around an obstacle. %
        Parameters in all simulations $\Lambda = 0.33$, $\phi = 0.14$, and $\gamma$ as labelled in (a). %
        Simulations performed in a square domain of length $L = 150\sigma$ featuring nine obstacles of diameter $R_\te{i}/\sigma = 16.6$ and $L_\te{obs} = 50\sigma$, and results averaged over each of those and over ten individual simulations each of length $T\Omega/(2\pi) = 300$. %
        Inset: Maximal veocity in profile against the substrate friction coefficient $\gamma$.
        (b)~Decay of the vortical order of flow created at the obstacle surfaces for varying friction coefficient $\gamma$. %
        (c)~Top panel: Maximum of at the contact value of $\Psi$ against friction. %
        Lower panel: Penetration depth of the radial order as in Fig.~\ref{fig:vortex_order}. %
        The values for the decay lengths corresponding to the nonzero friction coefficients are (in order of increasing $\gamma$) $\lambda/\sigma = 14.5,\, 10.8,\, 7.3,\, 3.2$. %
	}
	\label{fig:friction}
\end{figure*}

\subsection{Effect of substrate friction}
\noindent
Chiral active fluids are typically studied in two-dimensional active fluid sheets sedimented on a glass substrate~\cite{soni2019odd, bililign2022motile, mecke2023simultaneous} or at an interface~\cite{huang2021circular}. 
This implies that a finite friction $\gamma$ between the fluid layer and the substrate or the interface damps any created flow~\cite{mecke2024substrate}, an effect that can be related to a solvent damping length $\lambda$ which decreases with increasing friction.  %
Analytically this can be taken into account by considering a linear damping force density $-\gamma \bm{v}$ on the right-hand side of Eq.~\eqref{eq:incom-stokes}~\cite{boffetta2005effects,hosaka2021lift,alert2022active,hosaka2021nonreciprocal,daddi2025analytical, mecke2024emergent},  %
which allows us to characterize $\lambda\equiv\sqrt{(\nu + \nu_\te{R})/\gamma}$, with $\nu$ and $\nu_R$ the translation and rotational solvent kinematic viscosities (see Appendix for details).
In a MPC fluid, substrate friction has been introduced by incorporating a very small number of evenly scattered virtual particles in the collision step with zero-mean Maxwell-Boltzmann distributed velocities~\cite{mecke2024substrate}. The exchange of momentum between fluid and virtual particles serves as a momentum sink, such that changing the density of these virtual particles results in a change of the friction coefficient $\gamma$ and related solvent damping length $\lambda$.

In the case of a patterned environment, half the distance between the obstacles surfaces $d\equiv(L-D)/2$, where the average flows do necessarily cancel, determines the system dynamics together with the solvent damping length $\lambda$. 
For vanishing and small values of the friction, $\lambda \to \infty$ and in particular $\lambda > d$, the flows originating at the surface of neighbouring obstacles interact  while having opposite directions. Consequently, the flows created at the opposite surfaces of two neighbouring obstacles diminish each other. 
In this regime, increasing the friction decreases the interactions of the flows created at different obstacles at distances $d$, while the origin of the flow creation at the surface is still very strong. Hence, increasing the substrate friction diminishes the effect of neighbouring obstacle counter-flow and the overall flow increases. 
For larger values of the substrate friction, such that  $\lambda < d$,  the effect of the counter-flow becomes increasingly negligible and increasing the friction further results in the more intuitive decrease of the generated flow velocity, since increasing the friction ultimately decreases the colloids velocities. 

The azimuthal fluid velocity decays with the distance to the obstacle surface as shown in Fig.~\ref{fig:friction}a, with a clear non-monotonous dependence with the value of substrate friction, with a maximum for the $\lambda = d$ case. This is more clearly observed in the inset of Fig.~\ref{fig:friction}a, where the maximum of the normalised average colloid velocity near the obstacle boundary $v_\varphi(r_\ast)$ is depicted. The edge flow is maximised for $\lambda/d \approx 1$, a regime where the influence of the counter-flow over the length scale $d$ is damped, but the individual rotors actuation on the scale $\sigma$ is still large. %
For completeness, we also calculate the variation of the pinned vortex order parameter in Fig.~\ref{fig:friction}b which also shows a non-monotonous behaviour. This non-monotonocity has the same origin as the one for the decay of the azimuthal flow velocity, and can be very clearly observed in Fig.~\ref{fig:friction}c for the maximum pinned vortex order parameter~$\Psi_\te{max}$ and its penetration depth~$l^\ast$.

%################################################%
\section{Summary and conclusions}
\label{sec:summary}
%################################################%
\noindent
In this work, we study how a two-dimensional suspension of rotating colloids modifies its behavior due to the interaction with periodically arranged circular obstacles. 
A robust edge flux around each obstacle surface emerges in the direction predefined by the system's chirality, and coexists with the chiral active turbulent bulk behavior, this is the formation of multi-scale vortices. 
The relative importance of the systematic edge flow and the active turbulent behaviour is tuned by the edge flow strength and the fact that the obstacles additionally block the formation of large active turbulent eddies by virtue of the steric interactions between the rotors and the obstacles.
The edge mode decays with the distance to the obstacle surface, and its intensity can be maximised by modifying the rotor density, rotational velocity, the obstacle diameter, and also by the surface friction. 
These parameters also modify the coherence of the rotation around the obstacles. This is related to the intensity of the flow field and characterised by the here defined pinned vortex order parameter, which quantifies a smooth transition from chaotic to coherent vortex flows.
Specific to chiral active system in contact with circular obstacles is the appearance of transverse anti-symmetric and non-reciprocal interactions and unusual transport coefficients with anti-symmetric contributions to the diffusion tensor, which acts perpendicular to the direction of the density gradient and is absent in usual fluids. %
The diffusive dynamics is as well modified by the edge mode on the obstacle surface, since it promotes transport of rotors from one unit cell of the obstacle lattice to the next one by rotors rolling along the obstacle surface. 
The dependence with the obstacle size shows to be universal and an optimal tradeoff between transport facilitation and obstruction is found for a relative obstacle size of $\Lambda \simeq 0.1$.

Our results are not only of profound theoretical interest for the understanding of the transition from active turbulent dynamics to coherent vortex flows, but also carry implications for establishing design principles for transport in chiral active materials by introducing boundaries. 
When chiral active matter is under complex confinement conditions, the breaking of detailed balance in non-equilibrium systems can be directed to act in a predetermined direction~\cite{yang2016thermoosmotic}, such that steady-state currents and vorticity fields~\cite{nishiguchi2018engineering}, directed transport processes occur~\cite{medina2016cellular}, or apparent chaotic dynamics can be tamed~\cite{guillamat2017taming}. Accordingly, the study of chiral active matter in a patterned environment is an important field of research with promising applications for the design of synthetic smart materials with tailored behaviour.

\section*{Acknowledgements}\noindent
J.M. gratefully acknowledges the National Natural Science Foundation of
China for supporting this work within the Research Fund for International
Young Scientists under grant number 12350410368. J.M. and M.R.
gratefully acknowledge the Gauss Centre for Supercomputing e.V. (www.
gauss-centre.eu) for funding this project by providing computing time
through the John von Neumann Institute for Computing (NIC) on the GCS
Supercomputer JUWELS at Jülich Super-computing Centre (JSC) and the
Helmholtz Data Federation (HDF) for funding this work by providing services
and computing time on the HDF Cloud cluster at the Jülich Supercomputing
Centre (JSC). Y.G. acknowledges funding support from the Natural Science
Foundation of Guangdong Province [2024A1515011343].

\section*{Author declarations}

\subsection*{Conflict of interest}\noindent
The authors declare no competing interests.

\subsection*{Author contributions}\noindent
J.M. wrote the simulation code, performed the simulations and conducted the continuum-theoretical calculations. %
M.R. designed the investigation. %
J.M., Y.G., and M.R. discussed the results. %
M.R. and Y.G. provided supervisory oversight, resources, and critical editing of the manuscript.
J.M. and M.R. wrote the manuscript. %
All authors approved the final manuscript. %

\section*{Data availability}\noindent
The data that support the findings of this study are available from the
corresponding authors upon reasonable request.\\
The custom code for the simulations on GPUs is available from the 
corresponding authors upon reasonable request.

\section*{Appendix: Damping length}\noindent
The influence of an underlying substrate can be taken into account in the generalised Stokes equation by adding a linear friction term~\cite{daddi2025analytical, hosaka2021nonreciprocal, mecke2024substrate} to Eq.~\eqref{eq:incom-stokes} leading to
\begin{equation}
\label{eq:incom-stokes-friction}
     -\frac{1}{\rho} \nabla p_\te{eff} + \nu \nabla^2 \bm{v} + \nu\hat{e}_z \times \nabla (2\widetilde\Omega - \omega) - \gamma \bm{v} = 0\,,
\end{equation}
where $\nu = \eta/\rho$ and $\nu_\te{R} = \eta_\te{R}/\rho$
Moreover, transforming Eq.~\eqref{eq:incom-stokes-friction} to polar coordinates again, \eg, in order to describe the flow created around a rotating colloid, we obtain
\begin{align}
    0 = (\nu + \nu_\te{R}) \cb{\partial_r^2 v_\varphi + \frac{1}{r}\partial_r v_\varphi - \frac{v_\varphi}{r^2}} - \gamma v_\varphi\,.
\end{align}
This equation can be transformed to the Bessel equation and is solved by a superposition of Bessel functions of the first and second kind~\cite{mecke2024substrate}, decaying on the lengthscale $\lambda = \sqrt{(\nu + \nu_\te{R})/\gamma}$. %
Accordingly, in systems with non-negligible rotational viscosity, the characteristic decay length additionally depends on $\eta_\te{R}$, whereas in systems with $\eta_\te{R} \approx 0$ we obtain $\lambda = \sqrt{\nu/\gamma}$. %
In the MPC algorithm, the fluid particles' momentum transfer to the virtual substrate particles (momentum sink) can be calculated explicitly in terms of the simulation parameters~\cite{mecke2024substrate} and yields
\begin{align}
    \gamma = \frac{1}{h} \frac{\rho_\te{s}}{\rho + \rho_\te{s}} \pt{1 - \frac{1}{2(\rho - 1)}}\,,
\end{align}
where $\rho_\te{s}$ is the density of the virtual substrate particles.

\section*{References}\noindent
%\bibliography{Bibliography}

\end{document}